\documentclass[%
 aip,
 amsmath,amssymb,
 reprint,%
]{revtex4-1}

\usepackage{graphicx}
\usepackage{dcolumn}
\usepackage{bm}

\usepackage{mathptmx}
\usepackage{etoolbox}

\usepackage{xcolor}
\usepackage{floatrow}

\begin{document}

\preprint{AIP/123-QED}

\title{Gate tunable terahertz cyclotron emission from two-dimensional~\\ Dirac fermions}
\author{B.~Benhamou-Bui}
\author{C.~Consejo}
\author{S.~S.~Krishtopenko}
\author{M.~Szo{\l}a}
\affiliation{Laboratoire Charles Coulomb (L2C), UMR 5221 CNRS-Universit\'{e} de Montpellier, F-34095 Montpellier, France.}
\author{K. Maussang}
\affiliation{Institut d'Electronique et des Syst\`emes (IES),
UMR 5214 CNRS-Universit\'{e} de Montpellier, F-34000 Montpellier, France.}
\author{S.~Ruffenach}
\author{E.~Chauveau}
\affiliation{Laboratoire Charles Coulomb (L2C), UMR 5221 CNRS-Universit\'{e} de Montpellier, F-34095 Montpellier, France.}
\author{S.~Benlemqwanssa}
\affiliation{Institut d'Electronique et des Syst\`emes (IES),
UMR 5214 CNRS-Universit\'{e} de Montpellier, F-34000 Montpellier, France.}
\author{C. Bray}
\affiliation{Laboratoire Charles Coulomb (L2C), UMR 5221 CNRS-Universit\'{e} de Montpellier, F-34095 Montpellier, France.}
\author{X.~Baudry}
\author{P.~Ballet}
\affiliation{CEA, LETI, MINATEC Campus, DOPT, Grenoble, France.}
\author{S.~V.~Morozov}
\author{V.~I.~Gavrilenko}
\affiliation{Institute for Physics of Microstructures, Russian Academy of Sciences, GSP-105, 603950 Nizhny Novgorod, Russia.}
\affiliation{Lobachevsky State University of Nizhny Novgorod, 603022 Nizhny Novgorod, Russia.}
\author{N.~N.~Mikhailov}
\affiliation{A.V. Rzhanov Institute of Semiconductor Physics, Siberian Branch of Russian Academy of Sciences, 630090 Novosibirsk, Russia.}
\affiliation{Novosibirsk State University 630090 Novosibirsk, Russia.}
\author{S.~A.~Dvoretskii}
\affiliation{A.V. Rzhanov Institute of Semiconductor Physics, Siberian Branch of Russian Academy of Sciences, 630090 Novosibirsk, Russia.}
\affiliation{Tomsk State University, 634050 Tomsk, Russia.}
\author{B.~Jouault}
\affiliation{Laboratoire Charles Coulomb (L2C), UMR 5221 CNRS-Universit\'{e} de Montpellier, F-34095 Montpellier, France.}
\author{J.~Torres}
\affiliation{Institut d'Electronique et des Syst\`emes (IES),
UMR 5214 CNRS-Universit\'{e} de Montpellier, F-34000 Montpellier, France.}
\author{F.~Teppe}
\email[]{frederic.teppe@umontpellier.fr}
\affiliation{Laboratoire Charles Coulomb (L2C), UMR 5221 CNRS-Universit\'{e} de Montpellier, F-34095 Montpellier, France.}

\date{\today}

\begin{abstract}
Two-dimensional Dirac fermions in HgTe quantum wells close to the topological phase transition can generate significant cyclotron emission that is magnetic field tunable in the Terahertz (THz) frequency range. Due to their relativistic-like dynamics, their cyclotron mass is strongly dependent on their electron concentration in the quantum well, providing a second tunability lever and paving the way for a gate-tunable, permanent-magnet Landau laser. In this work, we demonstrate the proof-of-concept of such a back-gate tunable THz cyclotron emitter at fixed magnetic field. The emission frequency detected at 1.5~Tesla is centered on 2.2~THz and can already be electrically tuned over 250~GHz. With an optimized gate and a realistic permanent magnet of 1.0~Tesla, we estimate that the cyclotron emission could be continuously and rapidly tunable by the gate bias between 1 and 3~THz, that is to say on the less covered part of the THz gap.
\end{abstract}

\maketitle

\section{Introduction}

The ability to generate and manipulate THz waves opens up new avenues for exploring fundamental phenomena and developing innovative technologies in various fields such as wireless communication, imaging, spectroscopy, and sensing. Several techniques have already been developed among those Schottky \cite{Ward2004}, Gunn \cite{Heribert2004}, Resonant Tunneling Diode \cite{Asada2008}, Impact Ionizing Avalanch Transit Time diodes \cite{Mukherjee2007}, or quantum cascade lasers \cite{Faist1994,Williams2007}. Nonetheless, none of them are fully tunable over the entire THz gap \cite{Sirtori2002} and therefore a compact highly-tuneable source is still in great demand. In this context, the study of cyclotron emission has emerged as a promising area of research.

The idea of amplification and generation of electromagnetic waves at the frequency of cyclotron resonance (CR) arose in the late 1950s. At that time, the amplification of electromagnetic waves in the CR regime was obtained in plasma \cite{Twiss1958} and in vacuum electronics \cite{Pantell1959,Schneider1959}. On the other hand, by that time the cyclotron resonance had already become the standard method for studying the effective masses and band structure of semiconductors \cite{Dresselhaus1955,Dexter1956} that naturally yielded the discussion of theoretical possibility of stimulated cyclotron emission in semiconductors. Moreover, the small effective mass in bulk semiconductors gave hope to advance the radiation frequency deep into the THz range. Already early theoretical studies \cite{Tager1959,Lax1960} have shown that the non-linear dynamics of the charge carriers in the CR regime is the key ingredient for amplification and generation of electromagnetic radiation in semiconductors. Particularly, Lax \cite{Lax1960} was the first who demonstrated that the needed conditions can be achieved at the semiconductors with non-parabolic band dispersion resulting in non-equidistant Landau levels (LLs). The non-equidistance leads to the discrimination of optical transitions up and down from the filled LL resulting to the possibility of amplification and generation in the CR regime.

Although the ideas of vacuum and semiconductor LL lasers appeared simultaneously, their further development went in completely different ways. The active research of vacuum LL lasers yielded creating of gyrotrons \cite{Flyagin1977,Gaponov1981}, which still remain the most powerful sources of millimeter radiation nowadays, while investigations of the possibility of semiconductor LL laser realization were actually stopped in the early 1960s. The interest in semiconductor LL lasers resurfaced in the 1970s after development of the methods for numerical simulation of the scattering processes in electric and magnetic fields \cite{Alber1977,Shastin1980} revealed a number of new possibilities for inverted distribution of the charge carrier resulting in realization of light-hole LL laser \cite{Ivanov1983,Andronov1984,Komiyama1985,Unterrainer1990,Ivanov1992,Klimenko2009} and CR-based NEMAG (Negative Effective Mass Amplifier and Generator \cite{Andronov1982}) in bulk \emph{p}-Ge \cite{Andronov1986,Morimoto2009}. These two reported \emph{p}-Ge lasers remain the only ones semiconductor LL laser implemented so far. Given the hole mass in Germanium, to shift their operation frequencies into THz range unfortunately requires a substantial magnetic field. This problem, added to the need for an electric field close to the breakdown voltage of the semiconductor, have disqualified this technology in favor of quantum cascade lasers, which are yet poorly tunable in the THz range.

The dual need for a non-parabolic band structure and a very low electronic mass then naturally directs cyclotron emission research towards graphene and its relativistic massless Dirac fermions \cite{Morimoto2009,Wang2015,Wendler2015,Brem2017,Wendler2017,Brem2018}. However, it turns out to be still possible to find allowed optical transitions reactivating the Auger mechanism and no cyclotron emission has yet been observed in there. Alternatively, there are materials, which harbor relativistic-like particles other than Dirac fermions with the energy dispersion that can decrease or eliminate Auger recombination inherent in graphene. Particularly, as was demonstrated in 2019, HgCdTe bulk films, which host relativistic-like Kane fermions \cite{Orlita2014,Teppe2016,Krishtopenko2022} in the vicinity of topological phase transition, effectively overcome non-radiative Auger recombination and enables the observation of significant Landau emission \cite{But2019}.

Recently, Landau emission has been reported in HgTe quantum wells (QWs) close to the topological phase transition \cite{Gebert2023}. Given that their band structure mimics the one of 2D Dirac fermions \cite{Buttner2011,Marcinkiewicz2017,Kadykov2018}, the emission proves to be tunable not only by the magnetic field, but also by the density of carriers in the QW as measured across eight different samples with varying carrier concentrations. The observation of cyclotron emission from HgTe QWs was direct experimental evidence that additional quadratic terms in the dispersion of Dirac fermions suppress Auger recombination between Landau levels. 

Although these experimental findings indeed revive the interest to the Dirac materials in the context of tunable LL lasers, the tunability of Landau emission through the variation of carrier concentration in a single back-gated device has not yet been experimentally demonstrated. The aim of this work is therefore to verify experimentally a concept for voltage-tunable THz cyclotron emission from Dirac fermions in HgTe QWs. This indeed represents a significant milestone towards the development of a gate-tunable THz Landau laser at a fixed magnetic field operating with a permanent magnet.

\section{Material and model}

\begin{figure}
\includegraphics[scale=0.275]{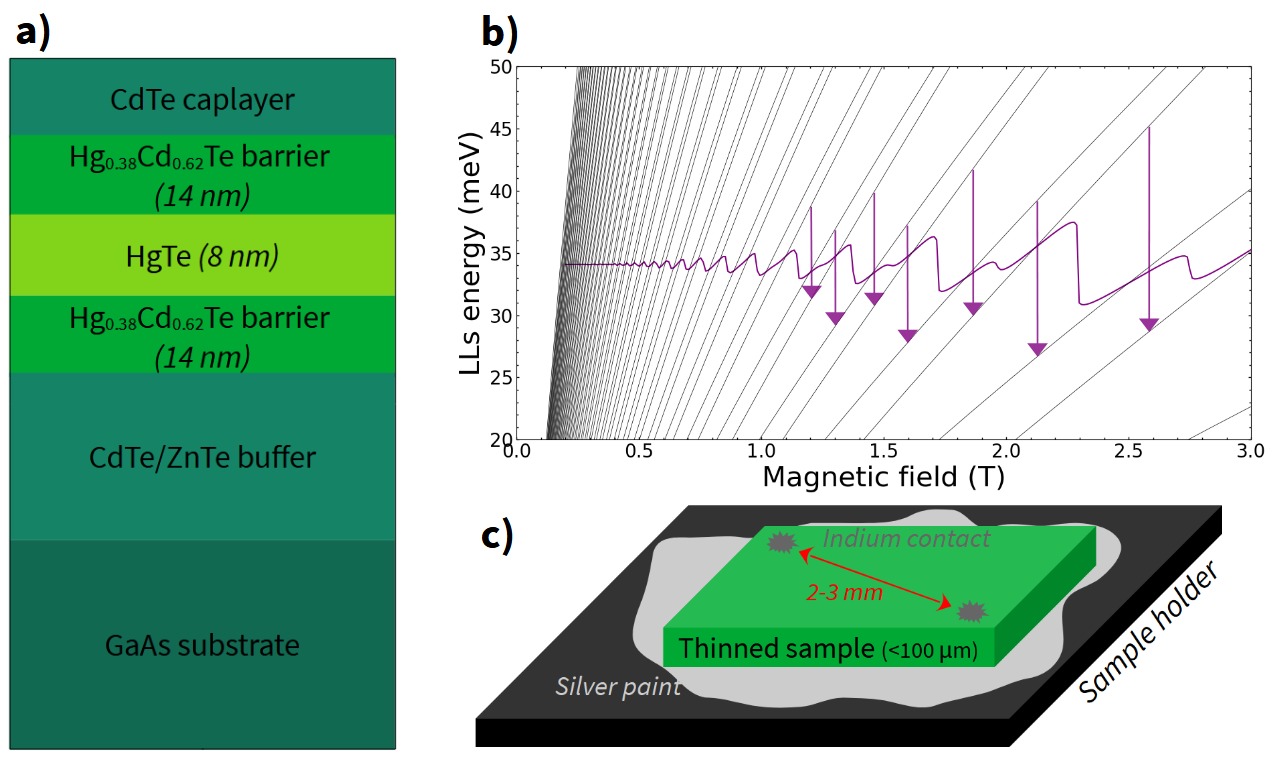}
\caption{\label{panel1} \textbf{(a)} Sketch of the sample's epitaxial growth. \textbf{(b)} Landau level fan chart calculated on the basis of the eight-band k$\cdot$p Hamiltonian (solid black lines). The evolution of the Fermi energy for an electron density of 3.35$\cdot 10^{11}$ $\mathrm{cm^{-2}}$ is represented by solid purple curve. The arrows schematically indicate the different inter-level transitions which contribute to the cyclotron emission. \textbf{(c)} Sketch of the sample resulting from the back-gate process.}
\end{figure}

The measurements were performed on a HgTe based QW grown by molecular beam epitaxy (MBE) on a (013)-oriented GaAs substrate, followed by a CdTe/ZnTe buffer to relax the lattice mismatch induced strain on the well \cite{Dvoretsky2010} (see Fig.~\ref{panel1}(a)). To interpret experimental results, we performed band structure and Landau level calculations based on the eight-band k$\cdot$p Hamiltonian for (013)-oriented heterostructures \cite{Krishtopenko2016}, which directly takes into account the interactions between $\Gamma_6$, $\Gamma_7$, and $\Gamma_8$ bands of bulk materials. In the calculations, we also take into account a tensile strain in the layers arising due to the mismatch of lattice constants in the CdTe buffer, HgTe QW, and Cd$_x$Hg$_{1-x}$Te barriers. The calculations were performed by expanding the envelope wave functions in the basis set of plane waves and by numerical solution of the eigenvalue problem. The energies of Landau levels (LLs) were found within a so-called axial approximation, while for the calculations of dispersion curves, non-axial terms were held. Details of calculations and the form of the Hamiltonian can be found elsewhere \cite{Krishtopenko2016}.

Figure~\ref{panel1}(b) shows a plot of the LLs fan chart resulting from this. The non-parabolicity in the energy dispersion is  linked to the non-equidistant spacing of LLs, which is a crucial requirement for the observation of cyclotron emission.
As discussed in \cite{Gebert2023} and observed in \cite{Orlita2012}, given the classical and quantum mobilities of our material, we are in a so-called ``incipient regime'', meaning in between a fully continuum of energy and a fully quantized regime. Indeed, in the range of magnetic field considered in this work, we are in a discrete LLs regime (see Fig.~\ref{panel1}(b)). But regarding the position of the Fermi level and the broadening of each level, a lot of LLs are involved in the cyclotron emission. In this regime, we already observe Shubnikov-de Haas oscillations but we can still treat the CR emission within the quasi-classical approach, where the photon energy is proportional to the magnetic field intensity:
\begin{equation}
    E=\dfrac{\hbar eB}{m_c}
    \label{CR}
\end{equation}
where $e$ is the electron charge, $B$ the magnetic field value, $\hbar$ the Planck constant and $m_c$ the cyclotron mass at the Fermi level. On the basis of the conduction band dispersion by applying the semiclassical quantization rule, it is possible to predict the behavior of this cyclotron mass with the electron concentration. This dependence allows to evaluate theoretically the tuning range of the resonant emission energy at given magnetic field with changing of electron concentration through the use of an electrical gate.

\begin{figure*}[t!]
\includegraphics[scale=0.38]{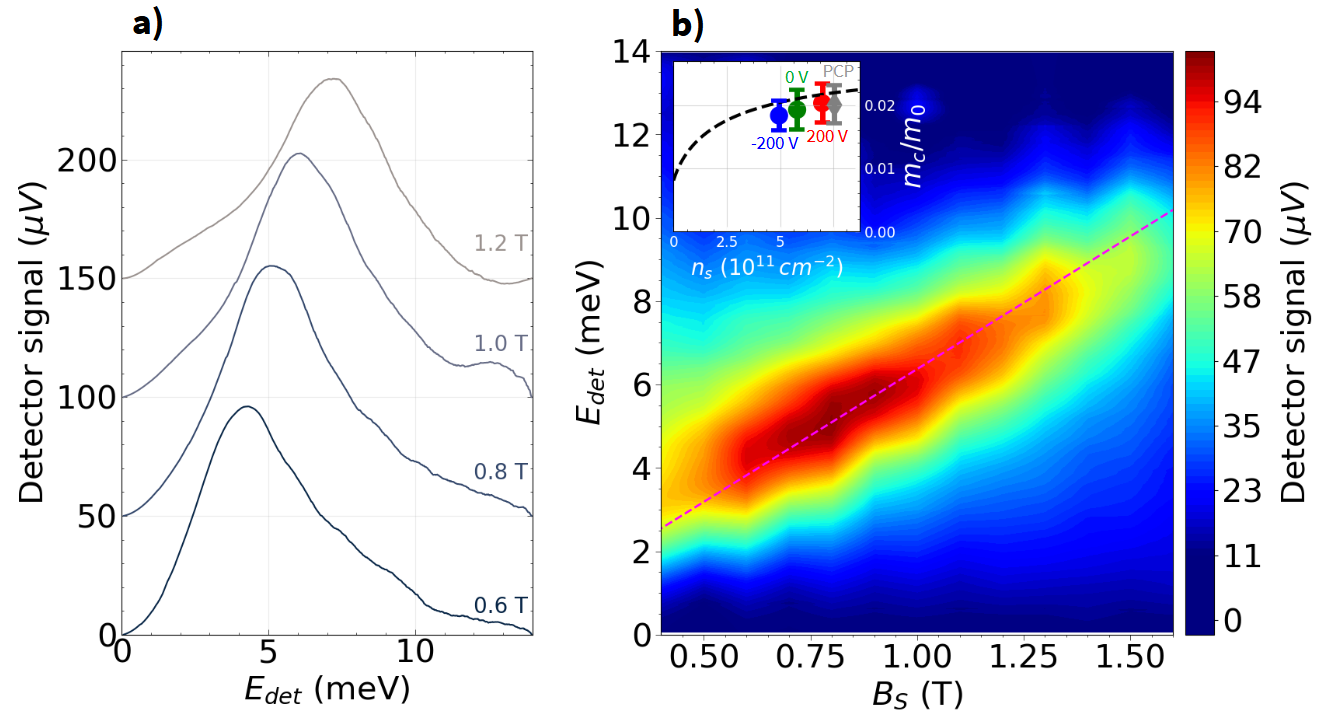}
\caption{\textbf{(a)} Waterfall plot of the electrical signal on the \emph{n}-InSb detector, witnessing an emission of photons, at zero gate voltage and for various magnetic field values. \textbf{(b)} Colormap of the evolution of the emission energy with the magnetic field applied on the sample. The dashed magenta line is a linear fit, based on Eq.~(\ref{CR}), allowing us to extract the cyclotron mass value. Experimental values of cyclotron mass for different electron density values are shown in insert. The value of the gate bias is indicated along with the experimental point. The predicted behavior, obtained within the eight-band k$\cdot$p Hamiltonian, is plotted in black dashed line. The PCP point is added in gray.}
\label{panel2}
\end{figure*}

\section{Experimental methods}

The primary concept \cite{Baenninger2012} is to utilize the substrate as a gate dielectric at low temperatures. To enhance the gate capacitance and optimize the gate efficiency, it is therefore crucial to thin down the substrate. The second step consists in gluing the thinned sample on a sample holder. This was done with conducting silver paint, allowing us to apply strong gate voltage values on the backside of the sample (see Fig.~\ref{panel1}(c)). In this work, voltage from -200 V to +200 V were applied. This range is limited by the breaking voltage thresholds of the substrate \cite{Baenninger2012}.

\begin{figure*}[hbt!]
\floatbox[{\capbeside\thisfloatsetup{capbesideposition={right,center},capbesidewidth=5cm}}]{figure}[\FBwidth]
{\caption{\textbf{(a)} Normalized emission spectra for various values of gate voltage, obtained at a fixed magnetic field value of 1.5~T. For the sake of clarity, not all spectra were represented. We added a spectrum obtained with PCP doping in gray. \textbf{(b)} Resonant frequency of emission lines as a function of the applied gate voltage value. The red dashed line is the best linear fit obtained and the shaded area corresponds to the 95$\%$ confidence interval.}\label{panel3}}
{\includegraphics[scale=0.37]{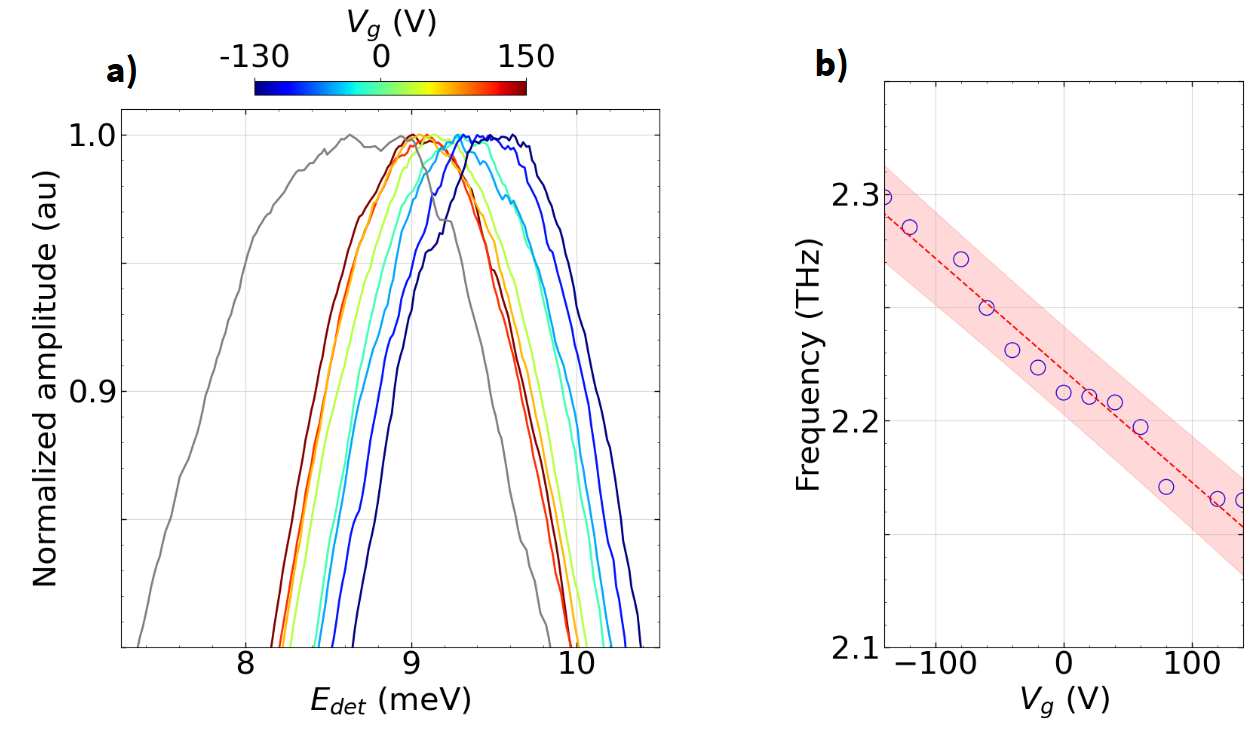}}
\end{figure*}

An alternative method of tuning the carrier concentration in the HgTe QW involves utilizing the persistent photoconductivity (PPC) effect, also known as ``optical gating''. This approach is often more convenient compared to fabricating gate structures \cite{Nikolaev2022}. The PPC phenomenon has been observed in various semiconductor materials, typically resulting in enhanced conductivity under illumination \cite{Spirin2018}. To enhance the electrical gating effect, we therefore employed this optical gating effect, which led to a notable augmentation in the carrier concentration during illumination. Consequently, we achieved an improved electrical gating action and a shift in the cyclotron frequency, as depicted in Fig.~\ref{panel3}(a).

The high energy LLs are populated thanks to short electrical pulses, of frequency 127~Hz and typical peak to peak amplitude of 8 V. The pulses are injected in the sample via Indium balls welded at the surface of the sample and which diffuse in the whole structure, enabling us to have ohmic contacts. The typical distance between two contacts is 2-3~mm, giving rise to electric fields in the sample of typically hundredth of $\mathrm{V\cdot cm^{-1}}$. We used a unique Landau spectrometer (see Supplementary materials of Ref.~\cite{Gebert2023} for the detailed experimental setup) integrated in a cryostat, along with three superconducting coils. Indeed, we use a hot-electron \emph{n}-InSb detector, which requires a strong magnetic field to narrow and tune its detection's energy window. Another coil is used to induce the LLs structuration in the sample and the last one allow us to compensate the influence of the other two coils on each other. Each measurement was performed at a constant magnetic field value on the sample, while scanning the magnetic field on the detector, thereby scanning the detection energy ($E_{det}$). The electrical signal on the detector is processed through a low-noise preamplifier and a lock-in amplifier. All the measurements were performed at 4.2~K.

\section{Experimental results}

Figure~\ref{panel2}(a) shows an example of the results that we obtained, for a gate voltage of $V_g=0$ V. One can see a waterfall plot of the electrical signal measured on the \emph{n}-InSb detector, each curve being recorded at a different magnetic field applied on the sample. Each curve features a peak at a given energy, witnessing a Landau emission coming out of the QW. It seems that this emission energy is growing with the magnetic field applied. It is also confirmed with Fig.~\ref{panel2}(b), which shows a colormap of the complete data-set obtained for this particular gate voltage value. One can clearly see that the emission energy evolves linearly with the magnetic filed applied on the sample, in accordance with Eq.~(\ref{CR}).

From a linear fit, we are able to extract the value of the cyclotron mass. The associated electron density, induced by the gate effect, has been obtained by the measurements of Shubnikov-de Haas oscillations~\cite{Ashcroft}. By performing such procedure for different values of back-gate voltage, on can follow the evolution of the emission peak and corresponding cyclotron mass values with the gate bias. Insert of Fig.~\ref{panel2}(b) shows the experimental values of cyclotron mass extracted for three extreme gate voltage values: -200 V, 0 V and +200 V. We can see that the experimental points are in good agreement with the prediction on the basis of the eight-band k$\cdot$p Hamiltonian (black dashed-line), as already observed in \cite{Ikonnikov2011}. This agreement is the confirmation that the light detected is really related to Landau-level emission in a HgTe QW. An extra experimental point, obtained by PCP doping, is added on the figure to show that is is possible to increase even more the accessible range of electron density in the sample.

As a proof of concept, we performed cyclotron emission measurements at a fixed low magnetic field of 1.5 T, for different values of the gate voltage applied. Figure~\ref{panel3}(a) shows the evolution of the emission peak's energy when we scan the gate bias from -130 V to +150 V. The change in energy is about 1 meV which corresponds to approximately 250 GHz. This is evidenced more clearly on Fig.~\ref{panel3}(b) where the peak maxima are extracted and plotted as a function of the gate voltage value. This results show a tunability of the emission frequency with the gate voltage value applied, around 2.2 THz and in a range of almost 200 GHz, which gives an electrical tunability of 485 $\pm$ 2 MHz/V. Again, we added a curve obtained by PCP doping (in gray in Fig.~\ref{panel3}(a)) to prove that it is possible to increase even more the emission's energy by adding more electrons in the QW.

\section{Discussion}
In comparison to \emph{p}-Ge, and without considering population inversion and laser action processes, HgTe QWs offer numerous advantages as cyclotron emitters. Firstly, due to the two-dimensional nature of the system, it inherently allows for the use of an electrical carrier density modulation gate, enabling the emission to be tuned at a fixed magnetic field and thus allowing the use of a permanent magnet. Second, the cyclotron mass of Dirac fermions in HgTe QWs, even with a small gap, is lower ($0.018{\cdot}m_0$ at zero gate bias) than that of holes in \emph{p}-Ge ($0.047{\cdot}m_0$)~\cite{Unterrainer1990}. This enables the generation of cyclotron emission within the THz range at significantly lower magnetic fields compared to those employed with \emph{p}-Ge (more than 3~T)~\cite{Unterrainer1990}.

\begin{figure}
    \centering
    \includegraphics[scale=0.33]{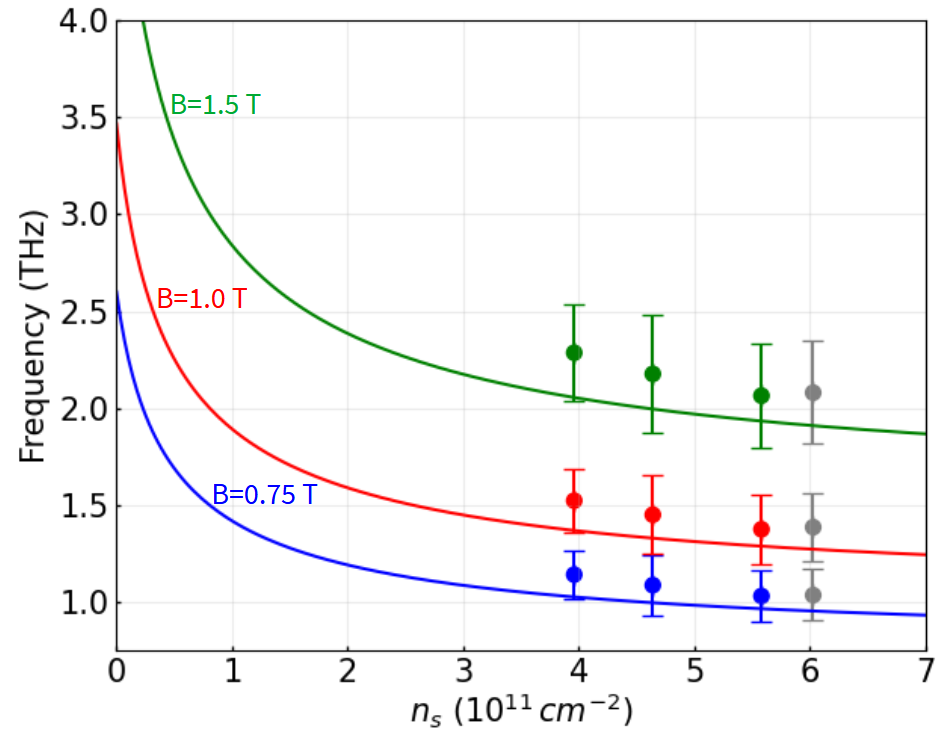}
    \caption{Predicted evolution of the cyclotron frequency with the electron density, for three different values of magnetic field (solid lines). The symbols represent expected resonant energies in given magnetic field evaluated on the basis of Eq.~(\ref{CR}) by means of the experimental cyclotron mass values, shown in Fig.~\ref{panel2}(c). The gray symbols correspond to the values obtained using the additional blue LED illumination.}
    \label{panel4}
\end{figure}

One interesting observation from the insert of Fig.~\ref{panel2}(b) is that the carrier concentrations we are working with are relatively high. According to the predictions on the basis of the eight-band k$\cdot$p Hamiltonian, the corresponding slope of the effective mass in this range is rather low. As illustrated in Fig.~\ref{panel4}, the frequency variation in relation to the carrier concentration demonstrates a nonlinear pattern that becomes more pronounced at lower doping levels. Consequently, it would be worthwhile to investigate lower electron density values in order to achieve more efficient tuning of the emission frequency with minimal carrier modulation.

Various approaches can be considered to accomplish this goal. One way is to enhance the gate efficiency from the backside through a more sophisticated technological process, possibly involving the use of an alternative dielectric material. Additionally, the gate efficiency can also be improved from the top by employing semi-transparent gates, typically used in magneto-transmission experiments \cite{Hughes2014,Muro1981}. Another option is to grow samples with lower doping levels, allowing us to operate directly within the highly non-linear range in which even a small change in carrier concentration leads to a significant variation in cyclotron mass and, consequently, in cyclotron frequency. In \cite{Baenninger2012}, it was reported that HgTe quantum wells with low doping can be grown by utilizing the effect of Hg vacancies.
If we consider the possibility of growing a QW with a zero gate bias density of around $2\cdot 10^{11}\, \mathrm{cm^{-2}}$, and assume the same level of gate efficiency demonstrated in this study (specifically, $\Delta n=1.5\cdot 10^{11}\,\mathrm{cm^{-2}}$ as shown in insert of figure (\ref{panel2})(b)), it would result in a much greater range of tunability in the emission frequency. Indeed, under these conditions, it would be possible to achieve a wide range of electron densities, ranging from $5\cdot 10^{10}\,\mathrm{cm^{-2}}$ to $3.5\cdot 10^{11}\,\mathrm{cm^{-2}}$. If we now report these concentrations on Fig.~\ref{panel4}, which shows the theoretical predictions in terms of emission frequency, we would observe a continuous tunability from $2.1$ THz to $3.5$ THz at a magnetic field value of 1.5 T. This would result in a tunability range of 4.5 GHz/V, which is nearly ten times greater than what we have demonstrated in this study.

In Fig.~\ref{panel4}, the theoretical tunability of a cyclotron source based on HgTe QWs is depicted for three distinct magnetic field values. The 0.75 T and 1 T values, achievable with a permanent magnet, provide realistic/practical insights into the frequency range in which such a compact source could effectively operate. It is worth emphasizing that despite the current limited tunability of the source, the underlying principle holds great potential. Such an emitter could be finely tuned to operate within the most challenging portion of the THz range, achieving remarkably high modulation speeds corresponding to the gate's modulation frequency. At a magnetic field strength of 1 T, the source's tunability could indeed span from 1 THz to 3 THz by varying the carrier concentration from $1\cdot10^{11}$ to $3\cdot10^{11}$ $\mathrm{cm^{-2}}$, enabling modulation frequencies in the GHz range.

\section{Conclusion and outlooks}
In summary, our study successfully demonstrated the proof of concept for a gate-tunable THz cyclotron source at a fixed magnetic field, by utilizing the Landau emission from 2D Dirac fermions in HgTe QWs through a straightforward technological process involving substrate thinning and a basic back-gate. Such a source holds the potential for modulation speeds reaching the GHz range, surpassing the limitations of magnetic field variation. Although the current tunability range is limited due to the simplicity of the technological process, our calculations indicate that it can easily cover the range from 1 to 3 THz by realistically achievable variations in carrier density using a high-quality gate. These findings pave the way for the development of a Landau laser capable of continuous tuning within the THz range, solely through modulation of an electrical gate voltage, while maintaining a fixed magnetic field provided by a permanent magnet.

\begin{acknowledgments}
This work was supported by the Terahertz Occitanie Platform, by the CNRS through IRP ``TeraMIR'' by the French Agence Nationale pour la Recherche for Equipex+ Hybat (ANR-21-ESRE-0026) and ``Colector'' (ANR-19-CE30-0032) projects, by the European Union through the Flag-Era JTC 2019 and by the Center of Excellence (Center of Photonics), funded by The Ministry of Science and Higher Education of the Russian Federation (contract no. 075-15-2022-316; S.V.M., V.I.G.). We would like to acknowledge Laurent Bonnet and Nassim Mouelhi for technical support.
\end{acknowledgments}

\nocite{*}
%

\end{document}